\documentclass[aps,prb,twocolumn,groupedaddress,floatfix,showpacs]{revtex4-1}
\usepackage{graphics,graphicx}
\usepackage{color}
\usepackage{amssymb,amsmath}
\usepackage{bm}
\usepackage{bookmath}

\newcommand{\req}[1]{Eq.~(\ref{#1})}

\newcommand{\reqs}[1]{Eqs.~(\ref{#1})}

\renewcommand{\Re}{{\rm Re}}

\begin{document}
\unitlength = 1mm
\title{Magnetic penetration depth in the presence of a spin-density wave in multiband superconductors at zero temperature}
\author{D. Kuzmanovski and M. G. Vavilov}
\affiliation{Department of Physics,
             University of Wisconsin, Madison, Wisconsin 53706, USA}

\date{\today}

\begin{abstract}
We present a theoretical description of the London penetration depth of a multi-band superconductor in the case when both superconducting and spin-density wave orders coexist. We focus on clean systems and zero temperature to emphasize the effect of the two competing orders. Our calculation shows that the supefluid density closely follows the evolution of the superconducting order parameter as doping is increased, saturating to a BCS value in the pure superconducting state. Furthermore, we predict a strong anisotropic in-pane penetration depth induced by the spin-density wave order.
\end{abstract}
\maketitle

\section{Introduction}
Most iron-based pnictide superconductors exhibit a phase diagram with the spin-density wave (SDW) and superconducting (SC) phases neighboring each other.\cite{Johnston2010,Stewart2011} More detailed studies of the boundary between the two phases indicated that the SDW and $s^{+-}$ SC order parameters\cite{Mazin2008,Chubukov2008} may coexist on a  microscopic level\cite{Chen2009,Nandi2010,Fernandes2010b,Laplace2009} due to the two-dimensional multi-orbital character of electronic band structure of these materials.\cite{Cvetkovic2009,Fernandes2010,Vorontsov2010} 
Measurements of nuclear magnetic resonance\cite{Laplace2009} and neutron scattering experiments\cite{Fernandes2010b} on electron-doped BaFe$_2$As$_2$ compounds support the presence of the coexistence phase between a pure superconducting and a pure SDW phase.  

The competition between the SDW and SC orders tends to suppress one order as the other order develops. 
Analysis of the relative weight of these two antagonistic order parameters reveals the microscopic nature of the coexistence region of the phase diagram, which can be assessed, for example, from thermodynamic characteristics of materials.  Indeed, as doping decreases and the material moves towards a stronger SDW phase, the specific heat jump at the onset of superconductivity decreases.\cite{Budko2009,Paglione2010,Johnston2010,Vavilov2011a,Vavilov2011}. Another thermodynamic quantity that shows a suppression of the SC order by the SDW ordering is the magnetic penetration depth.  Its measurements in iron-based superconductors is focus of many experiments (see references provided in Ref.~\onlinecite{Stewart2011}), and  recently were reported in the coexistence phase of Co-doped BaFe$_{2-x}$Co$_x$As$_2$ compounds.\cite{Gordon2010,Luan2011}

\begin{figure}[t]
\centerline{\includegraphics[width=\linewidth]{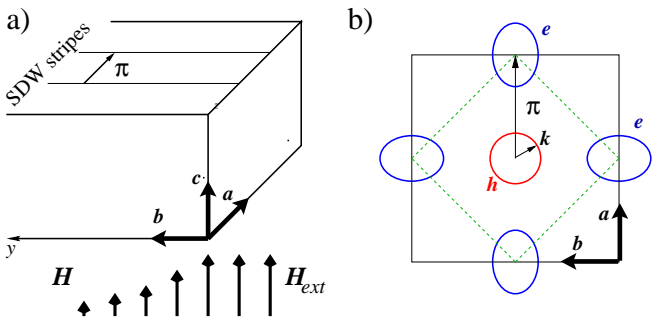}}
\caption{(Color online) a) Exponential decay of magnetic field. $(\bm{a},\bm{b},\bm{c})$ are the principal axes of the sample. The $1/e$ decay length is determined by the $Q_a$ principal value of the $Q$-tensor. $\bm{\pi}$ labels the SDW stripe order vector;
b) electronic structure of the multi-band model in the unfolded Brillouin zone. The hole FS is centered on the $\Gamma$ point, with SC order parameter $\Delta_{h}$, and the electron FS pockets are centered at $(\pm \pi, 0)$, $(0, \pm \pi)$, with SC order parameter $\Delta_{e}$. The magnetic order $m$ with momentum $\bm{\pi} = (\pi, 0)$ hybridizes hole and electron states separated by $\bm{\pi}$, but leaves FSs at $(0, \pm \pi)$ intact.
}
\label{fig:1}
\end{figure}

The London penetration depth $\lambda$ manifests itself as a length scale that determines the penetration of a \emph{static} magnetic field within the bulk of the superconductor. The magnetic field inside the superconductor is given by, see Fig.~\ref{fig:1}(a):
\begin{equation}
\label{eq:pendepth}
H(y)=H_{\rm ext}\exp\left(-y/\lambda\right).
\end{equation}
At zero temperature,
 $\lambda$ is defined by the superfluid density $n_s$
in the two-fluid model,\cite{Tinkham}
\begin{equation}
\label{eq:Q0}
\frac{1}{\lambda^2 }=  \frac{n_s q^2}{m^{\ast} c},
\end{equation}
where $q$ is the charge, and $m^{\ast}$ is the mass of the charge carriers in the superfluid phase. The BCS theory provides the same result, with the caveat that the charge carriers are Cooper pairs that carry charge $q = 2 e$, and have a mass $m{^\ast} = 2 m$. This result for $\lambda$ does not depend on the actual strength of the SC order parameter at zero temperature when $n_s$ is the total electron density. Because electron doping is not expected to drastically change the Fermi surface of BaFe$_{2-x}$Co$_x$As$_2$ compounds, the observed increase\cite{Gordon2010,Luan2011} of the magnetic penetration depth at zero temperature is likely due to the competition of superconductivity and SDW.

In this paper we present a microscopic theory of the magnetic penetration depth in the coexistence phase between SDW, and $s^{+-}$ SC states.\footnote{It had been shown\cite{Fernandes2010, Vorontsov2010} that an $s^{++}$ SC state does not coexist with SDW order.}
We utilize a simplified model for iron-based pnictides as two-dimensional multiband metal.  We focus on the zero-temperature limit when the magnetic penetration depth can be interpreted as the density of quasiparticles participating in formation of the SC order, while finite temperatures
lead to the reduction of the superfluid density, and increase in the London penetration depth.\cite{Tinkham} We also disregard the effects of impurities on the magnetic penetration depth. Therefore, we isolate the effect of the competition of SDW with the SC order parameter for the available electronic states from other known effects on the magnetic penetration depth.

Within the above assumptions, we obtained the magnetic penetration depth $\lambda$ in the coexistence phase. We show that $\lambda$ is determined solely by the density of quasi-particles related to the SC order parameter. Thus, as the doping moves the system from a pure SC phase into the coexistence phase, an SDW order develops and takes quasiparticles away from the SC condensate. As a consequence, 
$\lambda$ monotonically increases and becomes infinite at a transition to a pure SDW phase. This behavior is consistent with the results of measurements of the magnetic penetration depth in the coexistence phase.\cite{Gordon2010,Luan2011}

Our model also predicts an anisotropy of the magnetic penetration depth.  Earlier theoretical models\cite{Fernandes2010,Vorontsov2010}  predicted that the coexistence between homogeneous SC and commensurate SDW orders requires a weak anisotropy of an electron or hole Fermi surfaces (FSs). We demonstrate that even this weak anisotropy is sufficient to produce a significant \textit{in-plane} anisotropy of the magnetic penetration depth. 
At its peak value, the anisotropic part is about $30 \%$ of the value of the isotropic part. 
Unlike the isotropic part, this anisotropic part manifests itself within the coexistence region and vanishes in a pure SC phase. Measurements of the London penetration depth without taking into account the relative orientation of the crystal interface may lead to a large scattering of measured $\lambda$ in samples with identical doping, observed in Ref.~\onlinecite{Gordon2010}.

The outline of the rest of the text is as follows: Section II is a summary of the main results, together with a short description of the model within which we work. Section III contains the technical details of our calculations of the magnetic penetration depth in the coexistence phase. We present our conclusions in Section IV.

\section{Main Results}
We consider a multi-band model\cite{Fernandes2010,Vorontsov2010} of iron-based pnictide superconductors as an itinerant metal containing two electron, and one hole bands, see Fig.~\ref{fig:1}(b).  The FS for the hole band ($h$) is centered at the $\Gamma$-point $\{0,0\}$ of the unfolded Brillouin zone and the FSs of the electron bands ($e_{1,2}$) are centered at the $M$-points, defined by vectors $\bm{\pi}_1=\{\pm \pi, 0\}$ and $\bm{\pi}_2=\{0, \pm \pi\}$. 
At perfect nesting, the band dispersions satisfy $\xi_{e}(\bm{k}) = -\xi_{h}(\bm{k})$, where $\bm{k}$ is measured from the center of the corresponding band, and the two FSs are circular with equal sizes $k_F$ ($k_F \ll \pi/2$). 
A deviation from a perfect nesting may be tuned by doping  or pressure and  is parametrized by the small detuning parameter $\delta(\bm{k}) \equiv (\xi_{e}(\bm{k}) + \xi_{h}(\bm{k}))/2$.
 Another quantity that enters in later calculations is $\xi(\bm{k}) \equiv (\xi_{e}(\bm{k}) - \xi_{h}(\bm{k}))/2$. At detuning from perfect nesting, the hole pocket remains circular, whereas the electron pockets may change their size, as well as become elliptical. Both of theses effects can be accounted for by the expression
\begin{equation}
\label{eq:Deviation}
\delta_{\hat{\bm{k}}} = \delta_0 + \delta_2 \, \cos(2 \phi), \ \hat{\bm{k}} = \{ \cos(\phi), \sin(\phi) \}
\end{equation} 
with two band-structure parameters $\delta_{0,2}$.
We note that we need to consider $\vert \delta_{0,2} \vert\lesssim T_{s,0}  \ll v_{F} \, k_{F}$, where $v_F$ is the Fermi velocity of the fermions at the FS. 
We treat the two-body interactions in the SC and SDW channels in the self-consistent approximation by introducing SC, $\Delta$, and SDW, $m$ order parameters. We assume that the superconducting state has an extended $s^{+-}$-wave symmetry\cite{Mazin2008} and is characterized by an equal in magnitude, but opposite in sign isotropic SC order parameter on the FS, although a more general case, that also includes anisotropy of SC order parameter\cite{Chubukov2009,Maiti2011} deserves a separate consideration. The SDW order parameter is characterized by the nesting vector $\bm{\pi}_{1,2}$ that establishes an SDW stripe structure and breaks the $C_4$ crystal symmetry.\cite{Eremin2010} 
The self-consistency equations for $\Delta$ and $m$ are obtained from those derived in Ref.~\onlinecite{Vorontsov2010} for finite temperatures by taking the limit of $T \to 0$, as outlined in the end of Section III. The equations are\footnote{We disregard the other electron pockets on which no SDW, but only a SC order parameter develops. A pure SC order gives a constant contribution to the isotropic $Q$ parameter at zero temperature proportional to the density of states on that pocket. But, by neglecting this order parameter, we reduce the number of self-consistency gap equation from three to two and significantly simplify the analysis.}
\begin{subequations}\label{eq:scboth}
\be
\ln \frac{\vert \Delta \vert}{ \Delta_0 } = \int_{0}^{\infty}{d\omega \, 
\Re \left\langle \frac{ (E + i \delta_{\hvk})/E}{\sqrt{(E+i\delta_{\hvk})^2+\vert m \vert^2}}
-\frac{1}{E} \right\rangle},
\label{eq:scSCzerot}
\ee
\be
\ln \left( \frac{T_{c,0}}{T_{s,0}} \frac{\vert \Delta \vert}{  \Delta_0 } \right) = \int_{0}^{\infty}{d\omega \, \Re \left\langle \frac{1}{\sqrt{(E+i\delta_{\hvk})^2+ \vert m \vert^2}}
-\frac{1}{E} \right\rangle},
\label{eq:scSDWzerot}
\ee
\end{subequations}
where $E \equiv \sqrt{\omega^2 + \vert \Delta \vert^2}$, and $\langle \ldots \rangle$ stands for averaging over FS. Within this model, the SC transition temperature from a normal metal to a pure SC phase does not depend on detuning and is labeled by $T_{c,0}$, while the transition temperature from a normal metal to a pure SDW phase decreases as the detuning from perfect nesting increases. To characterize the interaction strength in the SDW channel, we introduce the transition temperature to a pure SDW phase at perfect nesting as $T_{s, 0}$. Parameter $\Delta_0$ in \reqs{eq:scboth} is the SC order parameter at zero temperature for a pure SC phase, $\Delta_0  = \pi e^{-\gamma_E} T_{c,0}$, with $\gamma_E \approx 0.577$ being the Euler constant.

The expressions for the response function $Q_{ik}$ that connects the supercurrent density and the vector potential as $j_i= -\sum _k Q_{ik}A_k$ are derived in the next section.  The tensor $Q_{ik}$ can be written as a combination of isotropic, $Q^{\rm iso}$, and anisotropic, $Q^{\rm an}$, components that are given by 
\begin{equation}
\left( \begin{array}{c}
Q^{\rm iso} \\ 
Q^{\rm an} 
\end{array} 
\right) = \frac{Q_0 \, {\Delta^2}}{{2}} \int\frac{ d\xi d\phi}{2\pi} 
\left( \begin{array}{c}
1 \\ 
\cos 2\phi
\end{array} 
\right)
\frac{\xi^2 Y(\xi)+ 4 m^2 Z(\xi)}{m^2+\xi^2} \label{eq:QzeroT}
\end{equation}
at the zero temperature limit. Here,
\begin{align}
Y & =\sum_\pm \frac{1}{2E_\pm^3},\quad Z=\frac{1}{2 E_+ E_- (E_++E_-)} 
\label{eq:Defs}\\
E_\pm(\xi,\phi) & =
\sqrt{(\sqrt{m^2+\xi^2} \pm \delta_{\hat{\bm{k}}} )^2 + \Delta^2}
\end{align}
 and $Q_0$ is given by 
the expression
\begin{equation}
\label{eq:QBCS1}
Q_0=\frac{2 e^2 v_F^2 N_F}{c}.
\end{equation}
in terms of the Fermi velocity, $v_F$, and the density of states at the Fermi energy, $N_F$, and does not depend on configuration of states far away from the Fermi surface.  

The electromagnetic response tensor $Q_{ik}$ has the principal axes that coincide with the crystallographic axes in directions $\bm{\pi}_{1,2}$.  If the crystal surface is perpendicular to the iron plane and is along one of these directions, the magnetic penetration depth have simple expressions 
\begin{equation}
\label{eq:minmax}
\frac{c}{4 \pi \, \lambda_{\rm min}^2}=Q^{\rm iso}+|Q^{\rm an}|, \quad
\frac{c}{4 \pi \, \lambda_{\rm max}^2}=Q^{\rm iso}-|Q^{\rm an}|.
\end{equation} 
In view of these simple relations between the magnetic penetration depth and the response tensor $Q$,
below we discuss the properties of the electromagnetic response tensor.

\begin{figure}[t]
\centerline{\includegraphics[width=\linewidth]{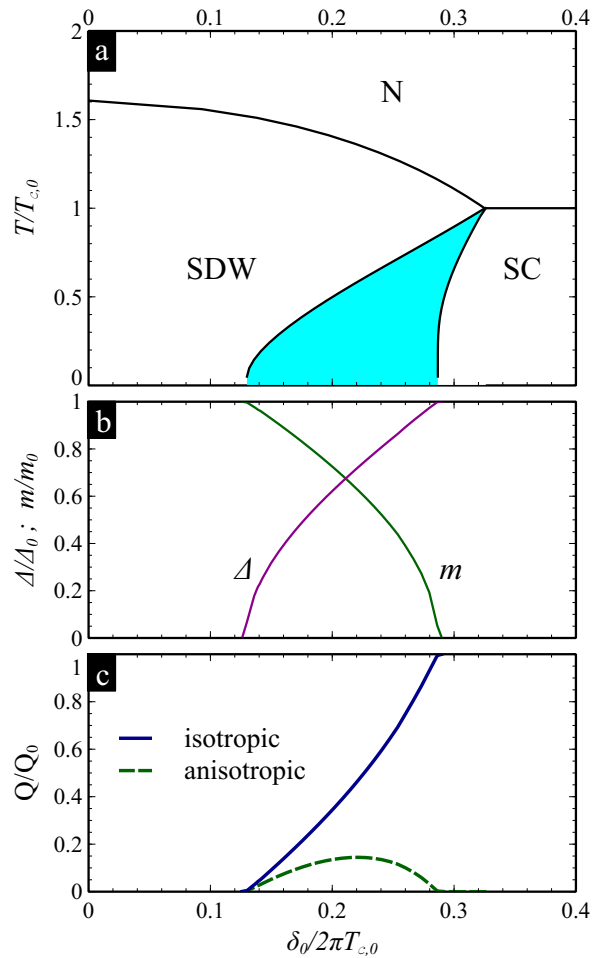}}
\caption{(Color online) a) The phase diagram of a two band metal with the interactions in SDW and SC channels as a function of the detuning of the FSs;
b) The zero-temperature SC and SDW order parameters as a function of detuning $\delta_0$ of the FSs; c) The isotropic (solid) and anisotropic (dashed) components of the $Q$-tensor as a function of detuning of the FSs. $T_{s,0}/T_{c,0} = 2.0$, and $\delta_2 = 0.4(2 \pi T_{c,0})$.
}
\label{fig:2}
\end{figure}

In Fig.~\ref{fig:2}, we present a case with parameters $T_{s,0}/T_{c,0} = 2.0$, and $\delta_2 = 0.4(2\pi T_{c,0})$. Panel (a) displays the phase diagram\cite{Vorontsov2010} as the isotropic detuning parameter $\delta_0$ is varied. The filled area represents the region of coexistence between SC and SDW orders. As the temperature is increased, the range of values for ${\delta_0}$ where coexistence is possible shrinks, degenerating into a tetra-critical point. We note that within this model, detuning plays a detrimental role only for the SDW order parameter. This is clear from the phase diagram, as we can see that the Neel temperature decreases. 
Panel (b) presents the evolution of the zero temperature order parameters $\Delta/\Delta_0$, and $m/m_0$, normalized to their values in pure SC, $\Delta_0$, or SDW, $m_0$, phases, respectively; note that $\Delta_0/T_{c,0} = m_0/T_{s,0}=\pi\exp(-\gamma_E)$.\cite{Vorontsov2010} As the detuning increases, the SC order parameter  monotonically increases from zero at the lower edge of the coexistence region to the full BCS value at the right edge of the coexistence region. The SDW order parameter follows an inverse trend. Using the values of the order parameters, the zero-temperature value for the response tensor is calculated according to \req{eq:QzeroT}. We note two limiting cases. First, when $\Delta = 0$, the integrand for both the isotropic and anisotropic component is a finite number, and, because of the prefactor $\Delta^2$, both 
$Q^{\rm iso}$ and $Q^{\rm an}$ vanish at $\Delta=0$. Second, when $m = 0$, the integrand decomposes into a sum for two independent bands $\pm \xi + \delta_{\hat{\bm{k}}}$. By a change of variable, the integrand can be made isotropic. Then, the integral for the isotropic component reduces to $1/\Delta^2$ for each band, and we obtain the value $Q_0$, \req{eq:QBCS1}. The average over the angles for the anisotropic component in Eq.~(\ref{eq:QzeroT}) is zero due to the $\cos(2\phi)$ term. These two limiting cases determine the behavior of the plots in panel (c). The isotropic component rises monotonically from zero to its BCS value, whereas the anisotropic component goes through a maximum in the coexistence region, and becomes zero at both edges of the coexistence phase.

\begin{figure}[t]
\centerline{\includegraphics[width=\linewidth]{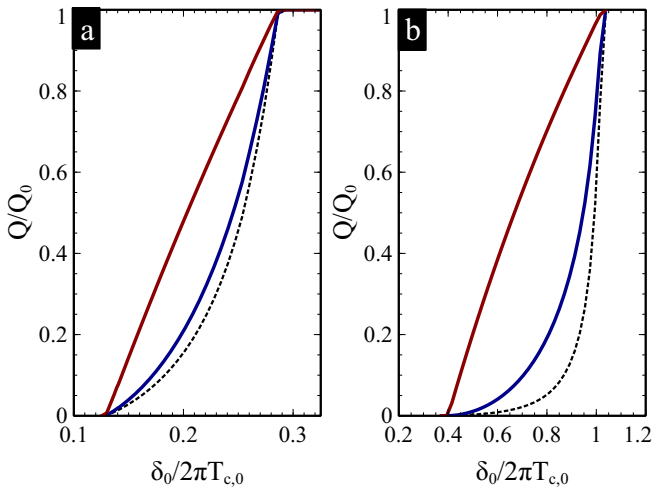}}
\caption{(Color online) The maximal and minimal values (solid) for the $Q$ parameter, and the value for no detuning, \req{eq:R} (dashed) for different values of the ratio $T_{s,0}/T_{c,0}$, a) $T_{s,0}/T_{c,0} = 2.0$. The coefficient for the anisotropic component of the detuning is $\delta_2 = 0.4(2 \pi T_{c,0})$; b) $T_{s,0}/T_{c,0} = 5.0$, and $\delta_2 = 1.0(2 \pi T_{c,0})$.}
\label{fig:3}
\end{figure}

\begin{figure}[t]
\centerline{\includegraphics[width=\linewidth]{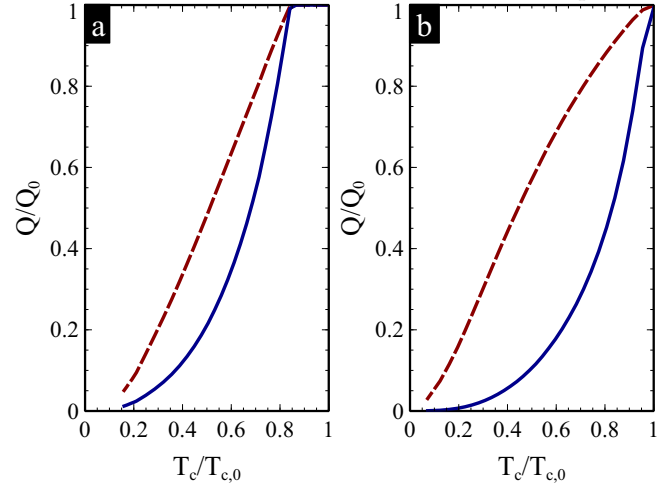}}
\caption{(Color online) A parametric plot of $Q_{\rm{max}}/Q_{0}$ (dashed) and $Q_{\rm{min}}/Q_{0}$ (solid) vs. $T_{c}/T_{c,0}$, as the detuning of the FSs is varied for different values of the ratio $T_{s,0}/T_{c,0}$  a) $T_{s,0}/T_{c,0} = 2.0$, and coefficient for the anisotropic component of the detuning $\delta_2 = 0.4(2 \pi T_{c,0})$; b) $T_{s,0}/T_{c,0} = 5.0$, and $\delta_2 = 1.0(2 \pi T_{c,0})$.
}
\label{fig:4}
\end{figure}

We present the minimal and maximal values for $Q$ in Figs.~\ref{fig:3} and \ref{fig:4} for two cases: a) $T_{s,0}/T_{c,0} = 2.0$, and $\delta_2/(2 \pi T_{c,0}) = 0.4$, and b) $T_{s,0}/T_{c,0} = 5.0$, and $\delta_2/(2 \pi T_{c,0}) = 1.0$. The difference between the plots is that in Fig.~\ref{fig:3}, $Q$ is presented as a function the dimensionless isotropic detuning parameter $\delta_0/(2 \pi T_{c, 0})$, whereas in Fig.~\ref{fig:4}, $Q$ is plotted versus the transition temperature, $T_c$, from a pure SDW phase to a co-existence phase, obtained for the same FSs detuning parameters $\delta_{0,2}$. $Q$ is normalized in units of $Q_0$, \req{eq:QBCS1}, and $T_c$ in units of $T_{c,0}$, which, in our model plays the role of the highest critical temperature reachable at optimal doping. Figure \ref{fig:4} attempts to eliminate model-dependent detuning parameters $\delta_{0,2}$, and express the dependence in terms of quantities that are available from direct measurements. In all cases, the presented dependence is a monotonic function. However, we observe that the anisotropy is relatively strong in the coexistence phase, which is manifested by a large spread of the bands formed by these plots. As the strength of magnetic coupling relative to SC coupling is increased, represented by the increase in the ratio $T_{s, 0}/T_{c, 0}$ from $2.0$ to $5.0$, the spread of these bands increases. This is in line with the notion that the anisotropy is caused by the SDW order, which scales as $T_{s,0}$.

The plots on Fig.~\ref{fig:4} become flat at the right edge of the coexistence region. This flat region is understandable if one considers the phase diagram in Fig.~\ref{fig:2}(a). Namely, the right end of the coexistence region shifts towards higher values of the detuning parameter ${\delta}_0$ as temperature increases. Thus, the tetracritical point takes place at higher detuning $\delta_0$ than the value of $\delta_0$ for transition from coexistence SDW+SC phase to a pure SC phase at zero temperature.  In this window of $\delta_0$, the response tensor is fixed, $Q = Q_{0}$, while $T_c$ still monotonically approaches $T_{c,0}$.  This gives rise to the flat part of the curve. Notice that this flat region is significantly reduced as ratio $T_{s,0}/T_{c,0}$ increases, since the above mentioned window of values for ${\delta}_0$ shrinks as well.\cite{Vorontsov2010}

Finally, we note that if we disregard the detuning parameter in expressions for the response functions, we arrive to the following simple expression:\cite{Fernandes2010a}
\begin{equation}
\label{eq:R}
Q^{\rm iso}=Q_0\frac{|\Delta|^2}{|\Delta|^2+|m|^2}.
\end{equation}
The co-existence between the SDW and SC order parameters requires a finite detuning with non-zero values of $\delta_0$ and $\delta_2$, and the detuning between the FSs has to be taken consistently both for evaluation of order parameters $\Delta$ and $m$, as well as for calculations of the response function $Q$.  Dashed lines in Fig.~\ref{fig:3} represent the result for $Q$ according to \req{eq:R} for order parameters evaluated for zero temperature limit.  This estimate for the response function is smaller than even the minimal value of the anisotropic response function obtained from \req{eq:QzeroT}.

\section{Calculation}
Here we work within a model developed in Ref.~\onlinecite{Vorontsov2010}. For definiteness, we consider the case of coexistence between an $s^{+-}$-wave SC order, and a commensurate SDW order. The system is defined through the mean-field Hamiltonian:
\begin{widetext}
\begin{equation}
\label{eq:Ham1}
\begin{array}{c}
\mathcal{H} = \frac{1}{2} \, \sum_{\bm{k} \alpha \beta}{\Psi^{\dagger}_{\bm{k} \alpha} \, (\hat{\mathcal{H}}_{\bm{k}})_{\alpha \beta} \, \Psi_{\bm{k} \beta}} + \sum_{n \in \lbrace h, e \rbrace, \bm{k}} {\xi_{n}(\bm{k})}, \\ \\
(\hat{\mathcal{H}}_{\bm{k}})_{\alpha \beta} = \left(\begin{array}{cc|cc}
\xi_{h}(\bm{k}) \delta_{\alpha \beta} & \Delta_{h} (i \hat{\sigma}_{2})_{\alpha \beta} & m^{\ast} (\hat{\sigma}_3)^{\dagger}_{\alpha \beta} & 0 \\

-\Delta^{\ast}_{h} \, (i \hat{\sigma}_{2})_{\alpha \beta} & -\xi_{h}(-\bm{k}) \delta_{\alpha \beta} & 0 & -m^{\ast} (\hat{\sigma}^\top_3)^{\dagger}_{\alpha \beta} \\
\hline
m (\hat{\sigma}_3)_{\alpha \beta} & 0 & \xi_{e}(\bm{k}) \delta_{\alpha \beta} & \Delta_{e} (i \hat{\sigma}_{2})_{\alpha \beta} \\

0 & -m (\hat{\sigma}^\top_3)_{\alpha \beta} & -\Delta^{\ast}_{e} (i \hat{\sigma}_{2})_{\alpha \beta} & -\xi_{e}(-\bm{k}) \delta_{\alpha \beta}
\end{array}\right)
\end{array},
\end{equation}
\end{widetext}
with $\Psi^{\dagger}_{\bm{k} \alpha} \equiv (c^{\dagger}_{h\bm{k} \alpha}, c_{h-\bm{k} \alpha}, c^{\dagger}_{e\bm{k} \alpha}, c_{e-\bm{k} \alpha})$ and $\Psi_{\bm{k} \alpha}$ being the conjugated column. $c_{e\bm{k}\alpha}$ and $c_{h\bm{k}\alpha}$ are electron and hole band annihilation operators, respectively. The two diagonal blocks of the matrix $(\hat{\mathcal{H}}_k)_{\alpha \beta}$ correspond to a purely SC system with $\Delta_{h}$ and $\Delta_{e}$ developing on two different bands, and two off-diagonal blocks contain SDW field $\bm{m} = m \, \hat{\bm{z}}$ that couples fermions between the two bands. In the case of $s^{+-}$-symmetry $\Delta_e = -\Delta_h = \Delta$.

If we consider a fictitious homogeneous vector potential $\bm{A}$, the only correction that we need to make is to substitute:
\begin{equation}
\label{eq:banddisp}
\xi_n (\bm{k}) \rightarrow \xi_n \left(\bm{k} - \frac{e}{c} \bm{A} \right) \approx \xi_{n}( \bm{k}) -\frac{e}{c} \, \bm{v}_{n\bm{k}} \cdot \bm{A} + \frac{e^2}{2 c^2} \, \bm{A} \cdot \stackrel{\leftrightarrow}{{m}}^{-1}_{n\bm{k}} \cdot \bm{A},
\end{equation}
Here, we introduced:
\begin{equation}
\label{eq:Blochvel}
\bm{v}_{n\bm{k}} \equiv \frac{\partial \xi_n}{\partial \bm{k}}(\bm{k})
\end{equation}
for the group velocity of Bloch electrons, and:
\begin{equation}
\label{eq:bandmass}
(m^{-1}_{n \bm{k}})_{i k} \equiv \frac{\partial^2 \xi_n}{\partial k_i \, \partial k_k} (\bm{k})
\end{equation}
for the components of the inverse band mass tensor.

The current density operator is obtained by differentiating Eq.~(\ref{eq:Ham1}) with respect to $\bm{A}$. Since we are interested in linear response of the current to the vector potential, it is sufficient to expand the Hamiltonian up to second order in powers of $\bm{A}$, as we did in Eq.~(\ref{eq:banddisp}). The expansion is straightforward, resulting in:
\begin{equation}
\label{eq:Current}
\begin{split}
\bm{J} & \equiv -c \, \frac{\delta H}{\delta \bm{A}} \\
& = \frac{1}{2} \, \sum_{\bm{k} \alpha \beta}{\Psi^{\dagger}_{\bm{k} \alpha} \, (\hat{\bm{J}}_{\bm{k}})_{\alpha \beta} \, \Psi_{\bm{k} \beta}} - \frac{e^{2}}{c} \sum_{n, \bm{k}}{\stackrel{\leftrightarrow}{m}^{-1}_{n\bm{k}} \cdot \bm{A}},
\end{split}
\end{equation}
where we introduced the following matrices:
\begin{align}
(\hat{\bm{J}}_{\bm{k}})_{\alpha \beta} & = e \, (\hat{\mathcal{\bm{V}}}_{\bm{k}})_{\alpha \beta} - \frac{e^{2}}{c} \, (\hat{\stackrel{\leftrightarrow}{\mathcal{M}}}^{-1}_{\bm{k}})_{\alpha \beta} \cdot \bm{A},  \label{eq:currentmat} \\
(\hat{\mathcal{\bm{V}}}_{\bm{k}})_{\alpha \beta} & = \delta_{\alpha \beta} \, \mathrm{diag}\left(\bm{v}_{h\bm{k}}, -\bm{v}_{h-\bm{k}}, \bm{v}_{e\bm{k}}, -\bm{v}_{e-\bm{k}} \right), \label{eq:velmat} \\
(\hat{\stackrel{\leftrightarrow}{\mathcal{M}}}^{-1}_{\bm{k}})_{\alpha \beta} & = \delta_{\alpha \beta} \, \mathrm{diag}\left(\stackrel{\leftrightarrow}{m}^{-1}_{h\bm{k}}, -\stackrel{\leftrightarrow}{m}^{-1}_{h-\bm{k}}, \stackrel{\leftrightarrow}{m}^{-1}_{e\bm{k}}, -\stackrel{\leftrightarrow}{m}^{-1}_{e-\bm{k}} \right). \label{eq:massmat}
\end{align}

The average of the single-particle operator Eq.~(\ref{eq:Current}) is expressible solely in terms of the single-particle Green function. However, in the mean-field approximation, the corrections to the Green's function itself are expressible in terms of the Green's function for the system described by Eq.~(\ref{eq:Ham1}):
\begin{equation}
\label{eq:Dyson}
(i \, \omega_{m} \hat{1} - \hat{\mathcal{H}}_{\bm{k}}) \, \hat{G}_{\bm{k}}(\omega_{m}) = \hat{1}.
\end{equation}
Expanding $\langle \bm{J} \rangle$ to linear order in powers of $A$, and identifying the coefficient of proportionality with the \textit{Q} response function:
\begin{equation}
\label{eq:Qparamdef}
\langle J \rangle_i = -Q_{i k} \, A_k,
\end{equation}
we arrive at the following expression:
\begin{equation}
\label{eq:QParam}
\begin{split}
Q_{i k} &= \frac{e^{2}}{2 c \beta} \sum_{\bm{k}, \omega_m}{e^{i \omega_{m} 0^{+}}\left[ \mathrm{Tr} \left(\hat{\mathcal{V}}_{i \bm{k}} \hat{G}_{\bm{k}}(\omega_{m}) \hat{\mathcal{V}}_{k \bm{k}} \hat{G}_{\bm{k}}(\omega_{m})\right) \right.} \\
 &+ \left. \mathrm{Tr}\left(\hat{\mathcal{M}}^{-1}_{i k \bm{k}} \hat{G}_{\bm{k}}(\omega_{m})\right) \right] + \sum_{n,\bm{k}}{(m^{-1}_{n\bm{k}})_{i k}}
\end{split}.
\end{equation}
At this point, we assume a general band dispersion relation, with only inversion symmetry $\xi_n(-\bm{k}) = \xi_n(\bm{k})$. According to Eq.~(\ref{eq:Blochvel}), and Eq.~(\ref{eq:bandmass}), we have $\bm{v}_{n-\bm{k}} = -\bm{v}_{n\bm{k}}$, and $\stackrel{\leftrightarrow}{m}^{-1}_{n-\bm{k}} = \stackrel{\leftrightarrow}{m}^{-1}_{n\bm{k}}$. The inverse matrix Eq.~(\ref{eq:Dyson}), with $\hat{\mathcal{H}}_{\bm{k}}$ given by Eq.~(\ref{eq:Ham1}) has a denominator of the form $(x^2 - E^2_{+}(\bm{k}))(x^2 - E^2_{-}(\bm{k})) \vert_{x = i \omega_m}$, where the excitation spectrum is given by:
\begin{equation}
\label{eq:ExcitationEnergiesMixed}
E_{\pm}(\bm{k}) = \sqrt{(\sqrt{\vert m \vert^2 + \xi^{2}_{\bm{k}}} \pm \delta_{\bm{k}})^2 + \vert \Delta \vert^2}.
\end{equation}
After taking the traces in Eq.~(\ref{eq:QParam}), we are left with rational algebraic expressions of $x^2$, and the partial fractional decomposition is made with respect to this variable. The sums of Matsubara frequencies are evaluated by using the following identities:
\begin{equation}
\label{eq:MatsSum1}
\frac{1}{\beta} \, \sum_{\omega_m}{\frac{e^{i \omega_m 0^{+}}}{(i \omega_m)^2 - E^2}} = \frac{n_F(E) - \frac{1}{2}}{E} \equiv g(E),
\end{equation}
\begin{equation}
\label{eq:MatsSum2}
\frac{1}{\beta} \, \sum_{\omega_m}{\frac{e^{i \omega_m 0^{+}}}{\left( (i \omega_m)^2 - E^2 \right)^2}} = \frac{g'(E)}{2 E},
\end{equation}
where $n_F(E) = (\exp (\beta \, E) + 1)^{-1}$ is the Fermi-Dirac distribution function.

Then, we combine the diamagnetic terms (containing factors of the inverse mass tensor) with the ``convective terms'' (containing products of two group velocities) by integrating by parts, through the use of the following identity which follows straightforwardly from the definitions Eq.~(\ref{eq:Blochvel}), and Eq.~(\ref{eq:bandmass}):
\begin{equation}
\label{eq:Bandrelation}
(m^{-1}_{n\bm{k}})_{i k} = \frac{1}{2} \left[ \frac{\partial v_{in\bm{k}}}{\partial k_{k}} +\frac{\partial v_{k n\bm{k}}}{\partial k_{i}} \right].
\end{equation}
After a simplification, we get the following expression:
\begin{equation}
\label{eq:QMixed}
Q_{i k} = \frac{e^2}{2 c} \, \sum_{\bm{k}}{\left[ L_{i k} \frac{g(E_{+}) - g(E_{-})}{E^{2}_{+} - E^{2}_{-}} + \sum_{\pm}{(K_{\pm})_{i k} \, g'(E_{\pm})} \right]},
\end{equation}
where, the coefficients $L_{i k}$, and $K_{i k}$ are given by:
\begin{subequations}
\label{eq:coeffs}
\be
L_{i k} = \frac{4 \, \vert m \vert^2 \, \vert \Delta \vert^2}{\vert m \vert^2 + \xi^2} (v_{e i} - v_{h i}) (v_{e k} - v_{h k}),
\ee
\be
\begin{split}
& (K_{\pm})_{i k} = \frac{\vert \Delta \vert^2}{(\vert m \vert^2 + \xi^2) \, E_{\pm}} \, \left[ v_{h i} v_{h k} \left( \sqrt{\vert m \vert^2 + \xi^2} \mp \xi \right)^2 \right. \\
& \left. + v_{e i} v_{e k} \left( \sqrt{ \vert m \vert^2 + \xi^2} \pm \xi \right)^2 + \left( v_{h i} v_{e k} + v_{e i} v_{h k} \right) \vert m \vert^2 \right].
\end{split}
\ee
\end{subequations}
Equations (\ref{eq:QMixed}) and (\ref{eq:coeffs}) can be simplified further if we assume a small deviation from the  perfect nesting and isotropic bands, defined by \req{eq:Deviation}. In this approximation, it is sufficient to assume $\bm{v}_{e\bm{k}} = -\bm{v}_{h\bm{k}} = v_{\bm{k}}$, just as in the case of perfect nesting and isotropic bands, which give a velocity in the radial direction $v_{\bm{k}} = v_{F} \, \hat{\bm{k}}, \ \hat{\bm{k}} = \{ \cos(\phi), \sin(\phi) \}$. Here, $\phi$ is measured from the direction specified by the nesting vector $\bm{Q}$. It may be shown that taking into account any deviation in the velocities is a small quantity of higher order. Under these approximations $L_{i k} = \left( 16 \vert m \vert ^2 \, \vert \Delta \vert^2/(\vert m \vert^2 + \xi^2) \right) \, v_{i} v_{k}$, and $(K_{\pm})_{i k} = \left( 4 \vert \Delta \vert^2 \xi^2/(\xi^2 + \vert m \vert^2) E_{\pm} \right) v_{i} v_{k}$. Next, in general, the symmetric two-fold 2D Cartesian tensor is reducible to an isotropic tensor (proportional to the Kronecker symbol $\delta_{i k}$), and a symmetric traceless tensor. However, one can show that under the above assumptions, the off-diagonal elements vanish identically. Indeed, a reflection relative to the y-axis, for example, changes the azimuthal angle as $\phi \rightarrow \pi - \phi$, and $\cos(2\phi) \rightarrow \cos(2\phi)$, thus the detuning parameter Eq.~(\ref{eq:Deviation}) remains unchanged. However, $x \rightarrow -x$, and $y \rightarrow y$, so $\langle x y f(\delta_{\hat{\bm{k}}}) \rangle \rightarrow - \langle x y f(\delta_{\hat{\bm{k}}}) \rangle$. Finally, we go from integration over $\bf{k}$ to integration over $\xi$, and take the slowly varying velocity and density of states in front of the integral as their values at FS. We present both the isotropic $Q^{\rm{iso}}$, and the anisotropic deviation $Q^{\rm{an}}$ (so that the principal values are $Q^{\rm{iso}} \pm Q^{\rm{an}}$):
\begin{widetext}
\begin{equation}
\left(\begin{array}{c}
Q^{\rm{iso}} \\

Q^{\rm{an}}
\end{array}\right) = \frac{e^2 v^2_F N_F}{c} \int_{-\infty}^{\infty}{d\xi \, \frac{\vert \Delta \vert^2}{\xi^2 + \vert m \vert^2} \, \left\langle \left(\begin{array}{c}
1 \\

\cos(2\phi) 
\end{array}\right) \, \left(\xi^2 \, \sum_{\pm}{\frac{g'(E_{\pm})}{E_{\pm}}} + 4 \vert m \vert^2 \, \frac{g(E_{+}) - g(E_{-})}{E^{2}_{+} - E^{2}_{-}} \right) \right\rangle} \label{eq:Qisoan},
\end{equation}
\end{widetext}
Finally, since we are mainly concerned with the pair-breaking effect the SDW order has on the Cooper pairs, as we vary the detuning parameters in Eq.~(\ref{eq:Deviation}), we take the $T \rightarrow 0$ limit in the above expressions. From Eq.~(\ref{eq:ExcitationEnergiesMixed}), Eq.~(\ref{eq:MatsSum1}), and Eq.~(\ref{eq:MatsSum2}), it follows that $g(E) \rightarrow -1/(2 E), g'(E) \rightarrow 1/(2 E^2)$ at zero temperature.

In order to calculate $Q$, one needs the values of the SC and SDW order parameters $\vert \Delta \vert$, and $\vert m \vert$. These are given by the self-consistency equations, that were derived elsewhere\cite{Vorontsov2010} for finite temperatures:
\begin{subequations}\label{eq:scbotht}
\be
\ln \frac{T}{T_{c,0}} = 2 \pi  T \sum_{\omega_m >0 }
\Re \left\langle \frac{ (E_m + i \delta_{\hvk})/E_m }{\sqrt{(E_m+i\delta_{\hvk})^2+\vert m \vert^2} }
-\frac{1}{\omega_m} \right\rangle \,,
\label{eq:scSCt}
\ee
\be
\ln \frac{T}{T_{s,0}} = 2 \pi T \sum_{\omega_m > 0}\;
\Re \left\langle \frac{1}{\sqrt{(E_m+i\delta_{\hvk})^2+ \vert m \vert^2} }
-\frac{1}{\omega_m} \right\rangle \,,
\label{eq:scSDWt}
\ee
\end{subequations}
where $E_m =  \sqrt{\omega_m^2 + \vert \Delta \vert^2 }$. We remind that  $T_{c,0}$ is the transition temperature for the pure SC state, and $T_{s,0}$ is the transition temperature for the pure SDW state at $\delta_{\hvk} =0$.

The above expressions are not defined at zero temperature. To reconcile with this, we add and subtract $1/E_m$ in each of the angle brackets. Then, the combination:
\[
2 \pi T \, \sum_{\omega_m > 0}{\left[ \frac{1}{E_m} - \frac{1}{\omega_m} \right]} = \ln \left( \frac{\pi e^{-\gamma_E} T}{\vert \Delta \vert} \right), T \rightarrow 0.
\]
We see that the logarithm of temperature cancels on both sides of \req{eq:scbotht}, and the remaining expressions have a well-defined zero temperature limit. Using this, and the BCS relation between the zero-temperature gap parameter and the critical SC temperature ($\vert \Delta_0 \vert = \pi e^{-\gamma_E} T_{c,0}$), we get the zero-temperature self-consistency conditions, presented above by \reqs{eq:scboth}.

\section{Conclusions}
To conclude, there is a clear reason behind the sensitive dependence of the magnetic penetration depth on doping  in the coexistence phase of SDW and SC orders in iron-based pnictide superconductors. In the presented model, doping is characterized by a detuning parameter of an electron and hole FSs from perfect nesting, which tends to suppress the SDW order in favor of SC order, eventually leading to a pure SC state. 
Although the FSs remain gapped due to both order parameters, only a fraction of quasiparticles near the FSs contribute to the superfluid density, while other quasiparticles participate in formation of the SDW. Therefore, the zero-temperature superfluid density increases, as doping is increased on the underdoped side. This behavior was recently observed in experiments.\cite{Gordon2010,Luan2011} 
Similarly, a previous attempt\cite{Vorontsov2009a} to fit experimental data on magnetic penetration depth temperature dependence in electron-doped $\mathrm{BaFe}_2\mathrm{As}_2$ required an adjustment for the zero-temperature value $\lambda_0$ of the magnetic penetration depth by factor of 4 between the optimally doped and underdoped samples. This adjustment in choice of $\lambda_0$ is consistent with the suppression of the superfluid density in the coexistence region.

One further prediction that our calculation gives is an anisotropic in-plane penetration depth, induced by the present SDW order. This would affect the measured value for the London penetration depth depending on the relative orientation of the sample boundary and the SDW nesting vector in mono-crystal samples. In multi-crystal samples, the anisotropy is likely to average out. 

We note that although the presented model captures the qualitative behavior, there is a need to consistently treat the effects of detuning\cite{Maiti2011} and disorder\cite{Vavilov2011} induced by doping not only on the competing orders, but also directly on the superfluid density. 

\acknowledgements
We are grateful to A. Chubukov, R. Prozorov and A. Vorontsov for fruitful discussions.
This work was supported by NSF Grant DMR 0955500, and the Wisconsin Alumni Research Foundation (WARF).



\end{document}